\documentclass[12pt,preprint]{aastex}
\usepackage{float,epsfig}
\usepackage{graphicx}
\usepackage{color}
\usepackage{natbib}
\usepackage[normalem]{ulem}
\newcommand{\degree}{^{\circ}}

\slugcomment{\color{blue} Version 16: 2014.10.20}

\begin{document}

\title{High-Time-Resolution Measurements of the Polarization of the Crab Pulsar at 1.38 GHz}


\author{
Agnieszka~S{\l}owikowska\altaffilmark{1},
Benjamin~W.~Stappers\altaffilmark{2},
Alice~K.~Harding\altaffilmark{3},
Stephen~L.~O'Dell\altaffilmark{4},
Ronald~F.~Elsner\altaffilmark{4},
Alexander~J.~van~der~Horst\altaffilmark{5},
Martin~C.~Weisskopf\altaffilmark{4}
}

\altaffiltext{1}
{Kepler Institute of Astronomy, University of Zielona G\'ora, Lubuska 2, 65-265
Zielona G\'ora, Poland}
\altaffiltext{2}
{Jodrell Bank Centre for Astrophysics, University of Manchester, Manchester M13 9PL, UK}
\altaffiltext{3}
{Astrophysics Science Division, NASA Goddard Space Flight Center, Greenbelt, MD 20771, USA}
\altaffiltext{4}
{Astrophysics Office, NASA Marshall Space Flight Center, ZP12, Huntsville, AL 35812, USA}
\altaffiltext{5}
{Astronomical Institute, University of Amsterdam, Science Park 904, 1098 XH Amsterdam, The Netherlands}




\begin{abstract}
Using the Westerbork Synthesis Radio Telescope (WSRT), we obtained high-time-resolution measurements of the full (linear and circular) polarization of the Crab pulsar. 
Taken at a resolution of 1/8192 of the 34-ms pulse period (i.e., $4.1~\mu{\rm s}$), the 1.38-GHz linear-polarization measurements are in general agreement with previous lower-time-resolution 1.4-GHz measurements of linear polarization in the main pulse (MP), in the interpulse (IP), and in the low-frequency component (LFC). 
We find the MP and IP to be linearly polarized at about $24\%$ and $21\%$, with no discernible difference in polarization position angle.
However, and contrary to theoretical expectations and measurements in the visible, we find no evidence for significant variation (sweep) in polarization position angle over the MP, the IP, or the LFC.
Although, the main pulse exhibits a small but statistically significant quadratic variation in the degree of linear polarization.
We discuss the  implications which appear to be in contradiction to theoretical expectations.
In addition, we detect weak circular polarization in the main pulse and interpulse, and strong ($\approx 20\%$) circular polarization in the low-frequency component, which also exhibits very strong ($\approx 98\%$) linear polarization at a position angle about $40\degree$ from that of the MP or IP.
The pulse-mean polarization properties are consistent with the LFC being a low-altitude component and the MP and IP being high-altitude caustic components.
Nevertheless, current models for the MP and IP emission do not readily account for the observed absence of pronounced polarization changes across the pulse.
Finally, we measure IP and LFC pulse phases relative to the MP that are consistent with recent measurements, which have shown that the phases of these pulse components are evolving with time.
\end{abstract}





\section{Introduction}

The Australia Telescope National Facility Pulsar Catalog \citep{Manc05} lists over 2300 radio pulsars.
Several radio studies \citep[e.g.,][]{Goul98,Kara06,Welt08} have measured the polarization for many of these pulsars.
Radio pulsars typically show moderate-to-strong linear polarization ($p_{L}$), being stronger
for those of higher spin-down energy-loss rate \citep[Figure~8]{Welt08}.
The linear polarization sometimes exhibits a characteristic swing or sweep of the position angle in an S-like shape near the pulse center, which is routinely interpreted in terms of the rotating vector model \citep[RVM,][]{Radh69}. 
For this model the point of emission is assumed to be in the polar cap region of the pulsar where a dipolar magnetic-field line points with a small angle (beamwidth) towards the observer.
The two free parameters of this simple model are the angle between the axes of rotation and the orientation of the magnetic dipole, and the view angle between the line of sight and the rotation axis. 
The variation of the radio position angle from some pulsars \citep[e.g.,][and references therein]{Lyne06} can be described by this model.

The Crab pulsar, the compact remnant of SN1054, and its pulsar wind nebula (PWN) are amongst the most intensively studied objects in the sky. 
The pulsar is one of the youngest and most energetic and its pulsed emission has been detected from 10 MHz \citep{Brid70} up to 400~GeV by VERITAS \citep{Aliu11} and by MAGIC \citep{Alek12}. 
The PWN is detected at energies up to 100~TeV \citep{Ahar04,Ahar06,Alle07,Abdo12}. 
Both the pulsar and nebula are predominantly sources of non-thermal radiation (synchrotron, curvature, and Compton processes), indicated not only by the broadband spectral continua but also by strong polarization in many wavelength bands \citep{Lyne06,Bueh13}. 

In the visible band, spatially-resolved polarimetry of the nebula, which began over a half century ago \citep{Oort56,Wolt57}, continues \citep[e.g.,][and references therein]{Mora13b}. 
Owing to its brightness, phase-resolved optical polarimetry of the pulsar has also been possible \citep{Jone81,Smit88,Slow09}.
However, phase-resolved X- and $\gamma$-ray polarimetry measurements of the Crab pulsar require space-based instruments, which have had limited sensitivity.
OSO-8 observations \citep{Silv78} of the Crab established only upper limits to the X-ray (2.6-keV and 5.2-keV) polarization of the pulsed emission. 
INTEGRAL IBIS observations \citep{Foro08,Mora13a} also detect no significant pulsed $\gamma$-ray (200--800-keV) polarization, although the off-pulse emission appears highly linearly polarized and is possibly associated with structures close to the pulsar rather than with the pulsar itself.

The Crab pulsar's light curve exhibits different features at different wavelengths, but it is currently the only pulsar for which the principal features persist over all wavelengths, from radio to $\gamma$-ray. 
There are two principal components---the main pulse (MP) and the interpulse (IP). 
This double-peak structure remains more-or-less phase-aligned over all spectral bands \citep{Moff96,Kuip01}.
One of several additional features in the radio band is the low-frequency component (LFC, e.g., \citeauthor{Moff96} \citeyear{Moff96,Moff99}),
having very low amplitude and occurring $\approx 0.10$ fractional pulse phase ($36\degree$) before the MP.
This component is most prominent around 1.4~GHz, in contrast with the ``precursor'' component \citep{Moff96}, which precedes the MP by $\approx 0.05$ fractional pulse phase ($19\degree$) at 0.327 and 0.610~GHz \citep[Table 2 of][]{Backer00}.

The MP and IP appear at roughly the same pulse phase from radio to $\gamma$-ray wavelengths, suggesting that their emission originates from a similar location in the magnetosphere at all wavebands.
Modeling of $\gamma$-ray light curves from the many pulsars observed by the {\sl Fermi Gamma-ray Space Telescope} \citep{Abdo13} strongly indicates that the high-energy emission originates in the outer magnetosphere, at altitudes comparable to the light-cylinder radius \citep{Roma10, Pier13, BaiSpit10}.
Outer magnetosphere emission models, such as the outer-gap \citep{Roma95}, slot-gap \citep{Musl04}, and current-sheet \citep{Petr05} had been proposed and studied prior to the {\sl Fermi} observations, but their emission geometry seems to account for the characteristics and variety of observed $\gamma$-ray light curves.
In addition, {\sl Fermi} has discovered a number of $\gamma$-ray millisecond pulsars whose radio peaks are nearly aligned with their $\gamma$-ray peaks \citep[e.g.,][]{Espi13}, like the Crab.
Modeling both $\gamma$-ray and radio light curves of these pulsars with the same outer magnetosphere emission models used to model young pulsars has suggested that their radio emission may originate from very high altitudes \citep{Vent12}.
Thus, in this paper we compare the phase-resolved radio polarization observations (\S\ref{s:obs}) that we have analyzed (\S\ref{s:res}) with such models (\S\ref{s:imp}).

\citet{Manc71} measured the linear polarization of the Crab pulsar's MP and precursor components at two radio frequencies.
The 0.410-GHz measurements found the MP to be $20\%$ linearly polarized at position angle $140\degree$ and the precursor to be $80\%$ linearly polarized at position angle $140\degree$. 
The 1.664-GHz measurements found the MP to be $24\%$ linearly polarized at position angle $60\degree$ and the precursor to be completely absent. 
As these measurements had rather large uncertainties and were obtained with a time resolution $1/256$ of the pulse period, they were quite limited for detecting variation of the linear polarization degree or position angle within a feature. 
However, \citeauthor{Manc71} noted a suggestion of rotation of the 1.664-GHz polarization position angle by about $30\degree$ through the MP.

More recently, \citet{Moff99} examined the pulse-profile morphology and polarization properties at three radio frequencies---1.424~GHz, 4.885~GHz, and 8.435~GHz---with a time resolution of $256~\mu{\rm s}$ (about $1/130$ of the pulse period).
The 1.424-GHz measurements found the MP to be $25\%$ linearly polarized at position angle $120\degree$; the IP, $20\%$ at position angle $120\degree$; and the LFC, $45\%$ at position angle $155\degree$.
\citeauthor{Moff99} note that the polarization position angle ``changes across the full period, although not significantly within components''.

Here we first report our observations (\S\ref{s:obs}), using the Westerbork Synthesis Radio Telescope (WSRT) in the Netherlands, of the full (linear and circular) 1.38-GHz polarization of the Crab pulsar, at high time resolution.
We then describe the polarimetry analysis and results (\S\ref{s:res} and Appendix~\ref{s:stat}) for the three pulse components, with a primary objective of determining the sweep of the position angle across each. 
Next we discuss the implications (\S\ref{s:imp}) of our measurements and analysis upon theoretical models for the pulsar emission.
Finally, we summarize our conclusions (\S\ref{s:sum}).

\section{The Observations} \label{s:obs}

The WSRT observations, on 2011 August 8, used 14 25-m-diameter dishes combined coherently to form the equivalent of a 94-m dish for pulsar observations. 
Owing to the interferometric nature of the WSRT, the observations partially resolve out the radio-bright Nebula, thus improving sensitivity over typical single-dish observations. 
Moreover, as the WSRT is an equatorially mounted telescope, there is no need to correct for parallactic angle.

\begin{figure}[ht]
\begin{center}
\includegraphics[angle=-90,width=0.9\columnwidth]{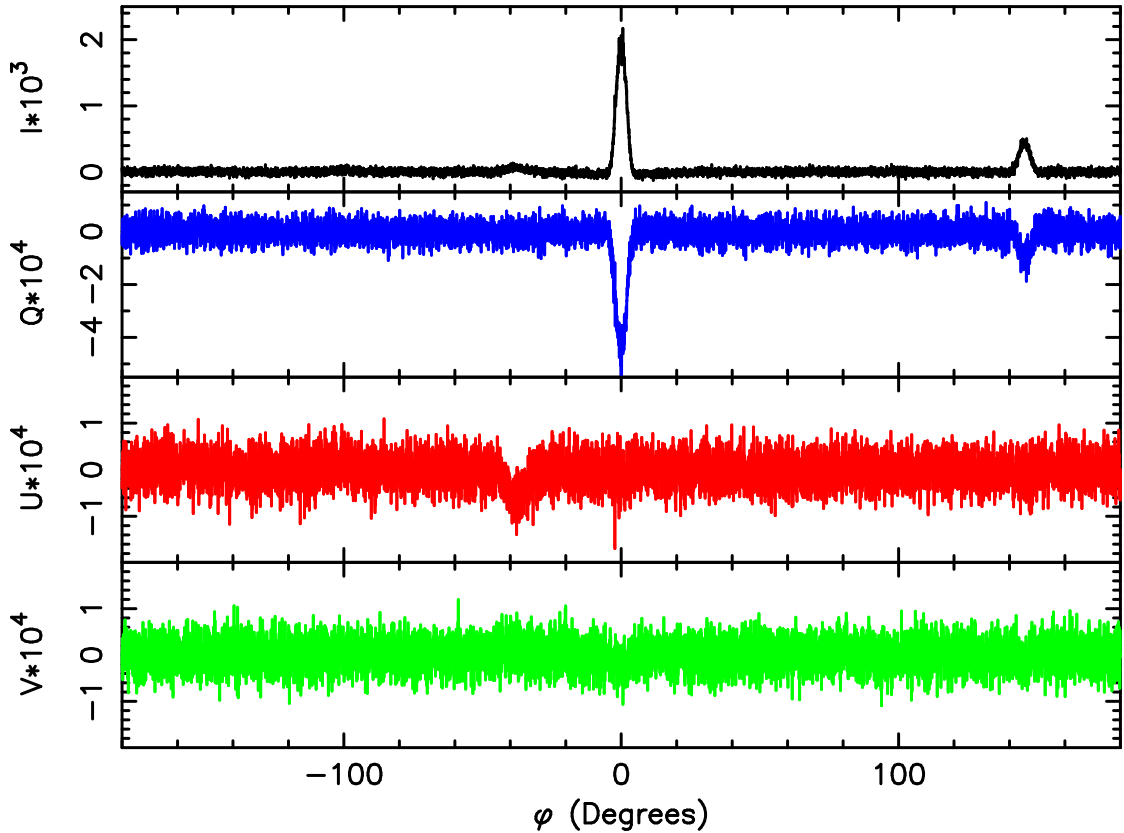}
\figcaption{Four Stokes parameters $I\times10^3$, $Q\times10^4$, $U\times10^4$, and $V\times10^4$ (arbitrary units) as functions of pulse phase $\varphi$, where the peak of the main pulse (MP) defines $\varphi = 0$.
The coordinate system for the Stokes parameters here sets $U = 0$ and $Q < 0$ for the MP.
\label{f:IQUV}}
\end{center}
\end{figure}

To combine coherently the dishes, correlated data from observation of a bright calibrator source is used to determine phase delays amongst dishes. 
This is accomplished using initially an unpolarized calibrator to determine delays between the two linear polarizations separately, followed by observation of a polarized calibrator to determine any residual delays between the two polarizations.
These procedures accurately calibrate the relative fluxes in the four Stokes parameters---hence, the polarization properties---but not the absolute flux.
Consequently, we express the Stokes measurements (e.g., Figure~\ref{f:IQUV}) in arbitrary units.

The PuMa-II \citep{Karu08} pulsar back-end was used to record Nyquist-sampled voltages at 8-bit resolution, across a 160-MHz band centered on 1380~MHz, for PSRs B0531+21 (Crab) and B0355+54, for a total of 144 and 18 minutes, respectively. 
The data were subsequently coherently de-dispersed and folded using the DSPSR \citep{vanS11} software package. 
Polarization profiles were formed after correcting for (frequency-dependent) interstellar Faraday rotation (rotation measure RM = $-42.3\pm 0.5~\rm{rad~m^{-2}}$) of the position angle, using the PSRCHIVE software package \citep{vanS12}. 
The polarization calibration was already carried out when forming the coherent sum of the dishes, nevertheless PSR B0355+54 was observed to verify that no further polarization calibration was required. 
Comparison with the profile observed by \citet{Goul98} showed that the polarization calibration matched exactly.
The Crab-pulsar profile was folded using the Jodrell Bank Ephemeris\footnote{http://www.jb.man.ac.uk/pulsar/crab.html}
with 8192 bins (about 4.1 $\mu$s/bin) across the pulse profile, matching the time resolution of the data after dividing into frequency channels and coherently de-dispersing. 
This time resolution was chosen also to match approximately the minimum broadening caused by scattering of the Crab pulsed emission by free
electrons in the Crab Nebula \citep[e.g.,][]{Backer00, Kuzmin08}. 

Figure~\ref{f:IQUV} displays our measurement of the four Stokes parameters $I$, $Q$, $U$, and $V$---which fully characterize the polarization---folded on the pulse period. 
Unfortunately, we were unable to determine the absolute polarization position angle for the Crab pulsar observation.
Instead, we selected a coordinate system for the Stokes parameters such that the MP has $U = 0$ and $Q < 0$.
Inspection of Figure~\ref{f:IQUV} immediately shows that our 1.38-GHz observations detect the flux and polarization of three components---MP, IP, and LFC.
Like the MP, the IP has $U \approx 0$ and $Q<0$; but the LFC has $U<0$ and $Q\approx 0$: 
Thus, the polarization position angles for the MP and the IP are roughly equal but differ from that of the LFC by about $40\degree$ (cf.~Eq.~\ref{e:psi}).
Similarly, but less obviously, the circular polarization of the MP and the IP are comparable, but that of the LFC has opposite polarity.

\section{Analysis and Results} \label{s:res}

The Stokes parameters have the virtues that they are statistically independent, typically exhibit Gaussian errors, and are directly superposable---i.e., each Stokes component ($I$, $Q$, $U$, or $V$) for multiple sources is the sum of the respective Stokes component for each source.
These properties follow from the fact that the Stokes parameters describe the polarization state in Cartesian-like coordinates.
This has the added virtue that there is no coordinate singularity at the origin, as occurs for polar-like coordinates---such as linear-polarization degree $p_{L}$ and position angle $\psi$.
Consequently, we perform all statistical analyses and model fitting (Appendix~\ref{s:stat}) on (pre-processed, Appendix~\ref{ss:proc}) raw Stokes data.

It is, of course, straightforward to transform to more customary parameters---e.g., linear-polarization degree $p_{L}$ (Eq.~\ref{e:pL}), position angle $\psi$ (Eq.~\ref{e:psi}), and  circular-polarization (signed) degree $p_{C}$ (Eq.~\ref{e:pC}):
\begin{equation}\label{e:pL}
p_{L}=\sqrt{(Q^2+U^2)}/I ;
\end{equation}
\begin{equation}\label{e:psi}
\psi=\frac{1}{2}\tan^{-1}(U/Q) ;
\end{equation}
\begin{equation}\label{e:pC}
p_{C}=V/I .
\end{equation}
\noindent For the three pulse features (MP, IP, and LFC), we estimate $p_{L}(\varphi_{n})$, $\psi(\varphi_{n})$, and $p_{C}(\varphi_{n})$ at each datum $n$ by substituting the measured $I_{n}$, $Q_{n}$, $U_{n}$, and $V_{n}$ into Equations~\ref{e:pL}, \ref{e:psi}, and \ref{e:pC}.

\begin{figure}[ht]
\begin{center}
\includegraphics[angle=-90,width=0.9\columnwidth]{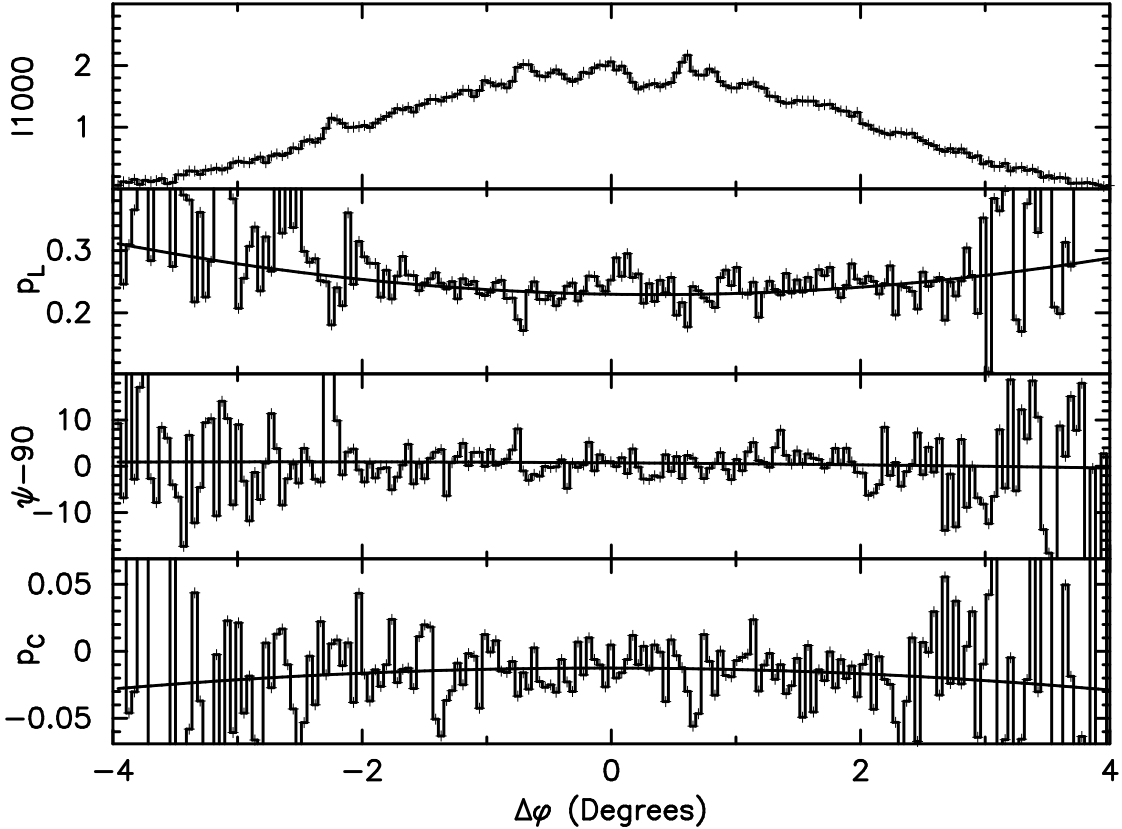}
\figcaption{Direct estimate of customary polarization parameters of the main pulse versus the phase-angle offset $\Delta\varphi$ from the MP center.
From the top, the plots display measured intensity $I$ data and then directly calculated fractional linear polarization $p_{L}$, position angle $\psi$, and fractional circular polarization $p_{C}$.
The smooth solid lines show the best-fit phase-dependent polarization properties based upon forward modeling of the Stokes data (Table~\ref{t:fit}).
\label{f:pol_MP}}
\end{center}
\end{figure}

\begin{figure}[ht]
\begin{center}
\includegraphics[angle=-90,width=0.9\columnwidth]{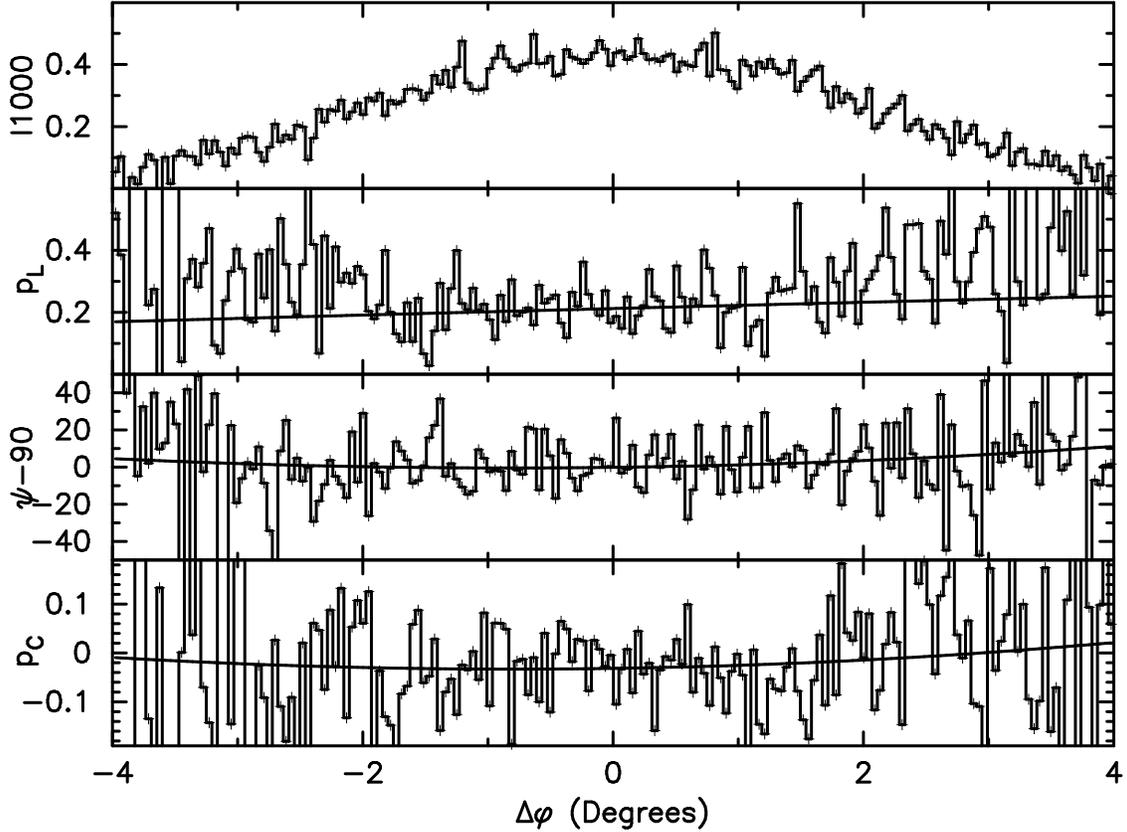}
\figcaption{Direct estimate of customary polarization parameters of the interpulse versus the phase-angle offset $\Delta\varphi$ from the IP center.
From the top, the plots display measured intensity $I$ data and then directly calculated fractional linear polarization $p_{L}$, position angle $\psi$, and fractional circular polarization $p_{C}$.
The smooth solid lines show the best-fit phase-dependent polarization properties based upon forward modeling of the Stokes data (Table~\ref{t:fit}).
\label{f:pol_IP}}
\end{center}
\end{figure}

Figures~\ref{f:pol_MP} and \ref{f:pol_IP} display the direct estimates of $I_{n}$, $p_{L\,n}$, $\psi_{n}$, and $p_{C\,n}$ over the MP and IP, respectively.
As the LFC is quite weak relative to the MP and the IP, the plots for the LFC are too noisy to display legibly.
Even for the stronger features---MP and IP---the RMS noise in the directly calculated polarization parameters ($p_{L\,n}$, $\psi_{n}$, and $p_{C\,n})$, which serves as an estimator of the statistical error, substantially increases away from the pulse center due to the low signal-to-noise ratio {\em per sample} in the pulse wings. 
In order to deal effectively with low-signal-to-noise data in the wings of the MP and IP and throughout the (weaker) LFC, we adopt a more rigorous forward-modeling approach to fit the measured Stokes data to the modeled $I(\varphi)$, $Q(\varphi)$, $U(\varphi)$, and $V(\varphi)$:

\begin{equation}\label{e:Q}
Q(\varphi) = I(\varphi) p_{L}(\varphi) \cos(2\psi(\varphi)) ;
\end{equation}
\begin{equation}\label{e:U}
U(\varphi) = I(\varphi) p_{L}(\varphi) \sin(2\psi(\varphi)) ;
\end{equation}
\begin{equation}\label{e:V}
V(\varphi) = I(\varphi) p_{C}(\varphi) .
\end{equation}

Appendix~\ref{s:stat} describes in some detail our approach for fitting polarization models to the Stokes data.
As Figures~\ref{f:pol_MP} and \ref{f:pol_IP} indicate that neither $p_{L}(\varphi)$, $\psi(\varphi)$, nor $p_{C}(\varphi)$ varies rapidly across the pulse profile, the approach simply models $p_{L}(\varphi)$, $\psi(\varphi)$, and $p_{C}(\varphi)$ as Taylor-series expansions in the phase-angle offset $\Delta\varphi \equiv (\varphi-\varphi_{0})$ from the center $\varphi_{0}$ of the respective pulse feature (MP, IP, or LFC).
Table~\ref{t:fit} tabulates the best-fit Taylor-expansion coefficients for the polarization dependence upon phase-angle offset:
\begin{equation}\label{e:pL0}
p_{L}(\varphi) = p_{L0} + p'_{L0} (\varphi-\varphi_{0}) + \frac{1}{2} p''_{L0} (\varphi-\varphi_{0})^2;
\end{equation}
\begin{equation}\label{e:psi0}
\psi(\varphi) = \psi_{0} + \psi'_{0} (\varphi-\varphi_{0}) + \frac{1}{2} \psi''_{0} (\varphi-\varphi_{0})^2;
\end{equation}
\begin{equation}\label{e:pC0}
p_{C}(\varphi) = p_{C0} + p'_{C0} (\varphi-\varphi_{0}) + \frac{1}{2} p''_{C0} (\varphi-\varphi_{0})^2.
\end{equation}

\begin{table}[ht]
\begin{center}
\caption {Best-fit polarization coefficients for the MP, IP, and LFC, using a single Gaussian for each pulse profile and up-to-quadratic variations in polarization functions $p_{L}(\varphi)$, $\psi(\varphi)$, and $p_{C}(\varphi)$. \label{t:fit}}
\begin{tabular}{ccccc} \\ \hline 
  Parameter              & Units                  & MP               & IP                 & LFC              \\ \hline
$\varphi_{0}-\varphi_{\rm MP}$ & $\degree$        & $\equiv 0$       & $145.389\pm 0.027$ & $-37.75\pm 0.19$ \\ 
$p_{L0}$                 & $\%$                   & $22.98\pm 0.30$  & $21.3\pm 1.0$      & $98.2\pm 6.7$    \\ 
$p'_{L0}$                & $\%/\degree$           & $-0.31\pm 0.19$  & $1.02\pm 0.62$     & $-0.8\pm 2.2$    \\ 
$p''_{L0}$               & $\%/\degree/\degree$   & $0.88\pm 0.22$   & $-0.02\pm 0.63$    & $0.0\pm 1.3$     \\ 
$\psi_{0}-\psi_{\rm MP}$ & $\degree{\rm PA}$      & $\equiv 0$       & $-0.1\pm 1.3$      & $40.8\pm 1.5$    \\ 
$\psi'_{0}$         & $\degree{\rm PA}/\degree$   & $-0.16\pm 0.20$  & $0.82\pm 0.78$     & $-0.16\pm 0.49$  \\ 
$\psi''_{0}$  & $\degree{\rm PA}/\degree/\degree$ & $-0.06\pm 0.21$  & $1.00\pm 0.89$     & $-0.21\pm 0.28$  \\ 
$p_{C0}$                 & $\%$                   & $-1.25\pm 0.20$  & $-3.15\pm 0.94$    & $20.5\pm 4.9$    \\ 
$p'_{C0}$                & $\%/\degree$           & $0.01\pm 0.13$   & $0.38\pm 0.56$     & $0.3\pm 1.7$     \\ 
$p''_{C0}$               & $\%/\degree/\degree$   & $-0.20\pm 0.15$  & $0.47\pm 0.57$     & $-0.49\pm 0.97$  \\ \hline
\end{tabular}
\end{center}
\end{table}

An important conclusion of this study is that the Stokes data are consistent---within statistical uncertainties---with constant polarization position angle $\psi$ across each of the three pulse features (MP, IP, and LFC) individually.
However, the MP does exhibit a small but statistically significant quadratic variation in the linear-polarization degree $p_{L}$.
While our 1.380-GHz polarimetry of the Crab pulsar has finer time resolution and better statistical accuracy than previous 1.424-GHz polarimetry \citep{Moff99}, measured values for the polarization degree and position angle (relative to MP) are mostly similar for the MP and for the IP.
The only significant difference is for the LFC's linear polarization degree and position angle.
We measured nearly total ($98\%\pm7\%$) linear polarization at a $+40.8\degree\pm 1.5\degree$ position-angle offset from the MP, 
whereas \citet{Moff99} found the LFC to be $\approx40\%$ linearly polarized at a $\approx +30\degree$ position-angle offset from the MP. 
We also detect circular polarization, which is moderately strong in the LFC ($20.5\pm 4.9\%$) but weak and opposite polarity in the MP ($-1.3\%\pm 0.2\%$) and in the IP ($-3.2\%\pm 0.9\%$).
In contrast with \citeauthor{Moff99}, we find no significant variation in the circular polarization across any of the three pulse components MP, IP, and LFC.

\begin{figure}[ht]
\begin{center}
\includegraphics[angle=0,width=0.9\columnwidth]{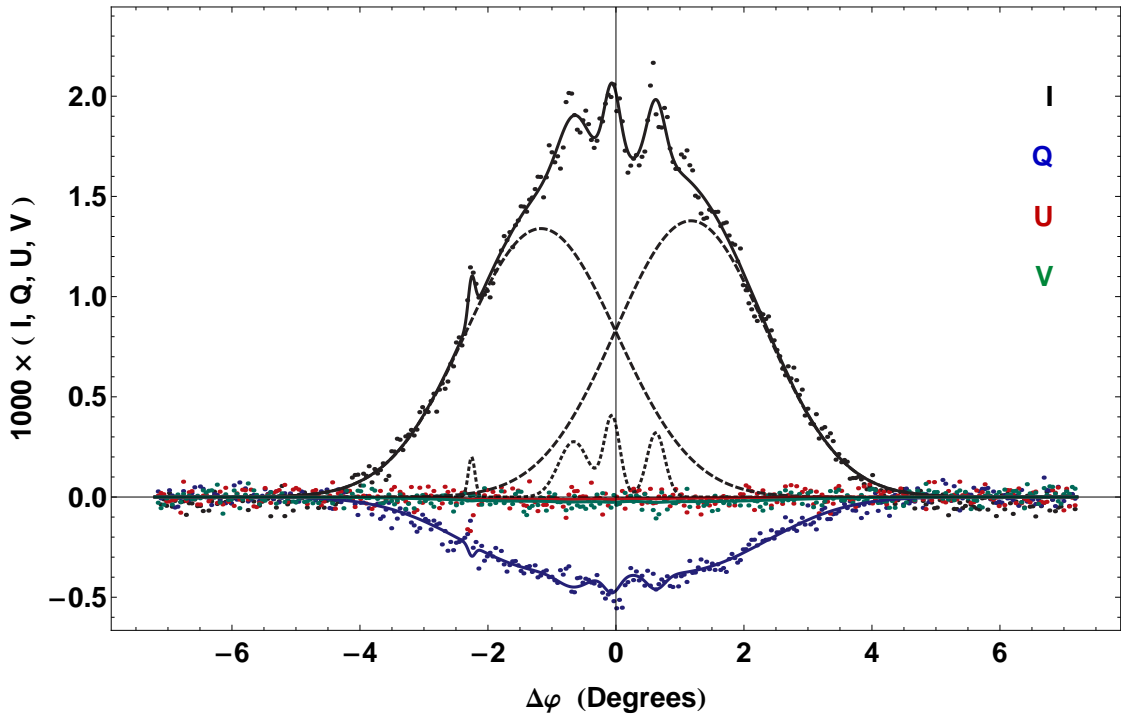}
\figcaption{Stokes data $I$, $Q$, $U$, and $V$ versus pulse phase offset $\Delta\varphi$ from the center of the main pulse (MP).
The lines represent the best-fit (minimum-$\chi^{2}$) Stokes functions for a multi-Gaussian profile and up-to-second-order variations in linear-polarization degree, position angle, and circular-polarization degree.
The pulse profile comprises 2 broad and 4 narrow Gaussians.
\label{f:IQUV_MP_2+4}}
\end{center}
\end{figure}

Another important conclusion---albeit peripheral to the polarimetry---relates to substructure in the pulse profile of the MP.
The fine time resolution and better statistical accuracy of our radio observation of the Crab pulsar resulted in measurement of statistically significant substructure (Appendix~\ref{ss:6-G}) in the profile of the main pulse (Figure~\ref{f:IQUV_MP_2+4}).
The typical width of the substructure is roughly $10\ \mu$s---i.e., $\le 0.1$ the width of the MP profile.
As the current analysis utilizes the sum of all data collected during the observation at a single epoch (2011 August 8), we have not assessed the temporal behavior of the profile.
However, we presume that this substructure results from sporadic, very strong giant radio pulses \citep{Bhat08,Karu10,Maji11,Hank03} occurring during the 144-minute observation.
Although the substructure is readily apparent in the $I$ profile of the MP, the discernible subpulses contribute only about $5\%$ of the fluence in the MP over the observation.
However, they likely result from only the strongest giant radio pulses in a distribution of pulse amplitudes.
Note that our conclusions as to the average pulse-phase dependences of the polarimetry are effectively independent of the precise modeling of the intensity profile of the MP.
On the other hand, inspection of the Stokes parameters (Figure~\ref{f:IQUV_MP_2+4}) or polarization parameters (Figure~\ref{f:pol_MP}) indicates that the polarization of some of the subpulses (e.g., at phase offset $\Delta\varphi\approx -2.3\degree$) differs substantially from the average polarization of the MP.

We also note that our WSRT-measured pulse-phase offsets of the IP and of the LFC from the MP are in good agreement with contemporaneous measurements at Jodrell Bank \citep{Lyne13}.
This tends to support the conclusion of \citet{Lyne13} that the phase separations of the IP and of the LFC from the MP are evolving with time.
Furthermore, the evolution of phase separations might contribute to the difference between our measurement of the LFC's polarization and earlier measurements \citep{Moff99}.

\section{Implications for Theoretical Models}  \label{s:imp}

Emission at altitudes comparable to the light-cylinder radius produce caustic peaks, formed by cancellation of phase differences due to aberration and retardation with that due to field-line curvature of radiation along the trailing magnetic-field lines \citep{Dyks03}.  
In outer-magnetosphere models, peaks in the light curves form when the observer's sight line sweeps across one or more bright caustic.
The caustics display distinct linear-polarization characteristics \citep{Dyks04}, including fast sweeps of  position angle and dips in polarization degree at the peaks, which are caused by piling up radiation emitted over a large range of altitudes and magnetic-field directions into the caustics. 
These characteristics are in fact seen in the optical polarization of the Crab pulsar \citep{Slow09}, which exhibits rapid swings of position angle across both the MP and IP, as well as dips in polarization degree to the 5\% level on the trailing edge of each peak.  

From the results presented in this paper, however, the characteristics of the radio linear polarization of the MP and IP resemble neither those of caustics in existing geometric models nor those observed in the optical emission.  
The lack of position-angle swing in the radio MP and IP is in stark contrast to the rapid position-angle swings in the optical.  
The very low circular polarization and moderate linear polarization observed here in the radio MP and IP are consistent with caustics, but the observed linear-polarization values ($\approx 22\%$) in the radio are significantly higher than those in the optical, and there is only a small variation with phase in the MP.  
On the other hand, the radio pulses are much narrower than the optical pulses, indicating that the radio MP and IP may originate along a smaller range of altitudes and/or in a subset of field lines.  

We have modeled the caustic emission and corresponding linear-polarization degree $p_{L}$ and position angle $\psi$ for the Crab pulsar, with a simulation using geometric renditions of standard slot-gap and outer-gap emission. 
These geometric emission models assume constant emissivity in the corotating frame along a set of field lines within the gaps, defined by a gap width $w$ across field lines in open-volume coordinates \citep{Dyks04}, where the width is a fraction of radius of open magnetic field lines. 
As in \citet{Dyks04}, the emission is assumed to occur over a fixed radius range, from minimum $r_{\rm min}$ to maximum $r_{\rm max}$.
For the simulations of Crab polarization here, we explored gap widths $w = 0.002, 0.01, 0.02, 0.05$, $r_{\rm min} = 0.3-0.9\,R_{\rm LC}$ and $r_{\rm max} = 0.5-1.2\,R_{\rm LC}$, where $R_{\rm LC} = c/\Omega$ is the light-cylinder radius. 
These are smaller ranges of altitude and smaller gap widths than in standard slot-gap or outer-gap models used in \citet{Dyks04}, which were $r_{\rm min} = R_{\rm NS}$, $r_{\rm max} = 0.95\,R_{\rm LC}$ for the slot gap and $r_{\rm min} = R_{\rm NC}$, $r_{\rm max} = 0.97\,R_{\rm LC}$ for the outer gap. 
Here $R_{\rm NS}$ is the neutron star radius and $R_{\rm NC}$ is the radius of the null-charge surface, at which the magnetospheric charge density in the corotating frame $\rho_{0} = \mathbf{\Omega \cdot B}/(2\pi c)$ vanishes.
  
We simulated emission using both retarded-vacuum-dipole \citep{Deut55}, as in \citet{Dyks04}, and force-free \citep{Cont10} magnetic-field geometries, as in \citet{Hard11}. 
Then we computed light curves and Stokes parameters for magnetic inclination angles $\alpha = 45\degree - 80\degree$, with $5\degree$ resolution for vacuum and $15\degree$ resolution for force-free magnetospheres, and observer viewing angles $\zeta = 55\degree-80\degree$ (both with respect to the rotation axis).
These ranges of $\alpha$ and $\zeta$ bracket the viewing angle of $60\degree-65\degree$ suggested by modeling of the X-ray torus \citep{Ng08}.
Following \citet{Dyks04}, \citet{Blas91}, and \citet{Hibs01}, we assume that the photon electric-field vector is parallel to the electron acceleration at each point along the field line to determine the Stokes parameters.

Although simulated light curves for the smaller gap widths produce narrower caustic peaks with less position-angle swing and depolarization, it is difficult to produce both $\psi(\varphi)$ and $p_{L}(\varphi)$ curves with no variation through the peaks.  
We compared a range of simulated light curves, $p_{L}$ and $\psi$ to the ones observed, and found that none of the models agree with the data.  
For the vacuum magnetospheres, the slot-gap model can produce appropriately narrow peaks for $w < 0.01$, but there is always some change in $\psi$ through both the MP and IP.  
At $\zeta = 60\degree$, there are dips in $p_{L}$ at only the first peak for $\alpha < 75\degree$ and dips at both peaks for $\alpha > 75\degree$.  
The outer-gap model produces a change in $\psi$ mostly in the IP but dips in $p_{L}$ in both peaks. 
While the force-free geometry, whose poloidal field lines are straighter than those in vacuum, can give a flatter position angle for certain inclination and viewing angles, the model's $p_{L}$ shows strong variation through the peaks in contradiction with the data.
For the force-free magnetospheres, the slot-gap model produces much less change in $\psi$ at the peaks for $\zeta = 55\degree-65\degree$ and $\alpha = 45\degree-75\degree$, but still not constant as observed.  
There is also a high level of depolarization in both peaks but $p_{L}$ is not constant through the peaks, as in the data.  
The outer gap in the force-free magnetosphere also produces changes in $\psi$ and $p_{L}$ in both peaks for these same ranges of $\alpha$ and $\zeta$.  

\begin{figure}[ht]
\begin{center}
\includegraphics[angle=-90,width=0.9\columnwidth]{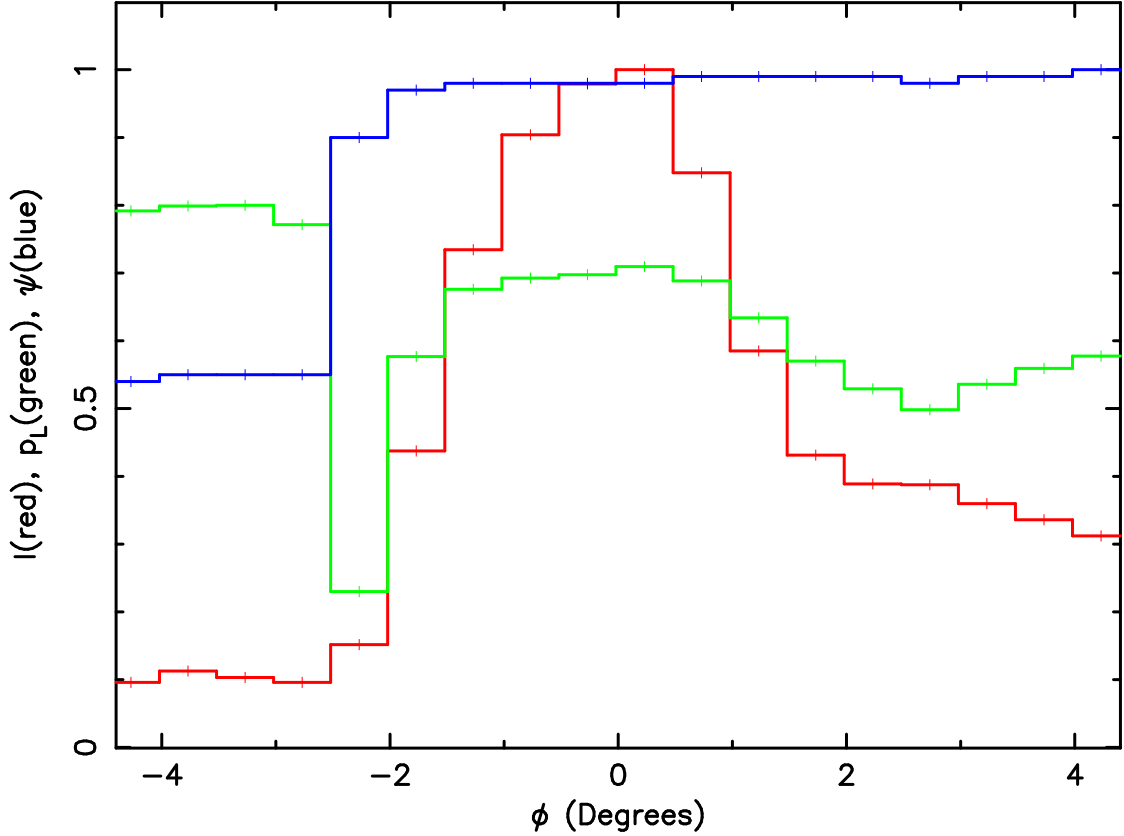}
\figcaption{Predicted relative variation through the MP of the intensity $I$ (red), linear polarization degree $p_{L}$ (green), and position angle $\psi$ (blue) for the slot-gap model, with a force-free magnetosphere.
For this case, the magnetic inclination angle $\alpha = 45\degree$ and observer viewing angle $\zeta = 60\degree$ with respect to the spin axis.
The ordinate range 0--1 corresponds to zero to peak intensity for $I$, 0\%--100\% polarization for $p_{L}$, and $-90\degree$ to $90\degree$ for $\psi$.
\label{f:fittheory}}
\end{center}
\end{figure}

For comparison with our measurement of the phase-resolved polarization properties of the Crab pulsar, we simulated 66 total (48 vacuum and 18 force-free) cases.
Based upon inspection of the results of these numerous simulated cases, the model light curve and polarization characteristics that seem to resemble most the Crab pulsar radio data is for the case of the slot-gap model in the force-free magnetosphere with $\alpha = 45\degree$ and $\zeta = 60\degree$.  
Figure \ref{f:fittheory} displays the results for this model for the MP. 
Note that this model does predict a rapid swing in polarization position angle and degree which we do not see; however, these swings occur on the preceding wing of the pulse, when the intensity is very low.

\begin{figure}[ht]
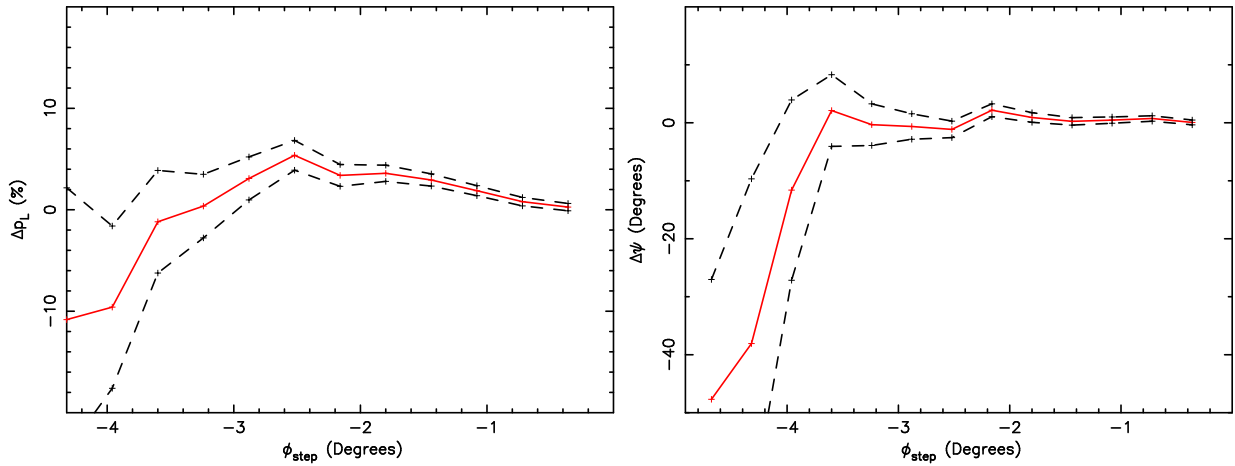

\begin{center}
\includegraphics[angle=-90,width=0.49\columnwidth]{fig_delP_vs_phase}
\includegraphics[angle=-90,width=0.49\columnwidth]{fig_delPA_vs_phase}
\figcaption{Constraints on a sharp step in the MP linear-polarization degree ($p_{L}(\varphi)$, left) and position angle ($\psi(\varphi)$, right) versus the putative step's pulse phase $\varphi_{\rm step}$ (relative to the MP center).
Large position-angle swings ($|\Delta \psi| > 45\degree$, say) are allowed (but not required) only very early ($\varphi_{\rm step} < -4\degree$) in the pulse---i.e., where the signal-to-noise ratio is low.
\label{f:delPPA}}
\end{center}
\end{figure}

In order to explore the possibility that the linear-polarization degree $p_{L}$ or position angle $\psi$ changes sharply in the preceding wing of the MP (as in Figure~\ref{f:fittheory}), 
we fit the Stokes data to a simple model of a step jump in the values of $p_{L}$ and of $\psi$ at a pulse phase $\varphi_{\rm step}$.
\begin{equation}\label{e:pLstep}
p_{L}(\varphi) = p_{L0} + \Delta p_{L}\; \Theta(\varphi_{\rm step}-\varphi);
\end{equation}
\begin{equation}\label{e:psistep}
\psi(\varphi) = \psi_{0} + \Delta \psi\; \Theta(\varphi_{\rm step}-\varphi).
\end{equation}
Here, $p_{L0}$ and $\psi_{0}$ are the best-fit values for constant linear-polarization degree and position angle; $\Delta p_{L}$ and $\Delta \psi$, the pre-step differences in the value of each; and $\Theta(\varphi_{\rm step}-\varphi)$, the unit step distribution ($=1$ for $\varphi<\varphi_{\rm step}$, 0 otherwise).
Figure~\ref{f:delPPA} shows the best-fit differences and their (1-sigma) uncertainties as functions of pulse phase of the step (relative to pulse center).
From this analysis, we conclude that any position-angle swing must be small---$|\Delta \psi| < 10\degree$ for $\varphi_{\rm step} > -3.5\degree$.
A large position-angle swing---$|\Delta \psi| > 45\degree$, say---is consistent with the data (but not required) only for $\varphi_{\rm step} < -4\degree$.
Note that the analysis requires $\Delta p_{L}>0$ for $\varphi_{\rm step} \ge -2.5\degree$ (and allows it for earlier $\varphi_{\rm step}$), as  this analysis does not include the small positive second derivative $p''_{L0}$ in the linear-polarization degree, which the Taylor-expansion fit to the MP Stokes data requires (cf.\ Table~\ref{t:fit}).

It is possible that the radio linear polarization in the MP and LP is very sensitive to the magnetic-field structure.
Existing models explored only the two extremes of vacuum (accelerating fields but no plasma) and force-free (plasma but no accelerating fields), neither of which describe real pulsars.  
More realistic, dissipative magnetosphere models with finite conductivity now exist \citep{Kala12,Li12} and should be used to model light curves and polarization characteristics.  
It is also possible that the radio emission in the MP and IP occurs along sets of field lines that lie deeper within the open/closed field boundary or the current sheet and have different polarization properties.

The low-frequency component (LFC) is substantially weaker than the MP and IP at 1.4~GHz.
As its name suggests, the LFC is not detected at radio frequencies higher than a few GHz and has no corresponding component in the visible band.
The nearly complete radio polarization ($p_{L}\approx 98\%$ and $p_{C}\approx 20\%$) of the LFC support the hypothesis that it is a highly coherent, low-altitude component. 
Note that the (lower frequency) precursor is also believed to be a highly coherent, low-altitude component, due to its high polarization and steep spectrum \citep{Rank90}.

\section{Conclusions} \label{s:sum}

Our 1.38-GHz observations of the Crab pulsar measured significant linear and circular polarization in the three most prominent pulse components---the main pulse (MP), inter pulse (IP), and low-frequency component (LFC).
These results are mostly in agreement with previous measurements of linear polarization at similar radio frequencies \citep[cf.][]{Moff99}.
The MP and IP are moderately linearly polarized ($p_{L}\approx 23\%$ and $21\%$, respectively) at the same position angle ($\psi_{\rm IP}-\psi_{\rm MP}\approx 0$); they are weakly circularly polarized ($p_{C}\approx -1.3\%$ and $-3.2\%$, respectively).
In contrast, the LFC is very strongly linearly polarized ($p_{L}\approx 98\%$), at a position angle $+40\degree$ from that of the MP or IP, and moderately circularly polarized ($p_{C}\approx 20\%$).

The fine time resolution (Period/8192 = 4.1 $\mu$s) and good sensitivity of the measurements at the Westerbork Synthesis Radio Telescope (WSRT) enabled a meaningful search for changes in linear-polarization degree $p_{L}$, in position angle $\psi$, and in circular-polarization degree $p_{C}$ across each of the three pulse components.
Neither the MP, IP, nor LFC exhibits a statistically significant change in the polarization position angle or circular polarization across the pulse.
For the MP, the linear term (``sweep'') is well constrained: $\psi'_{0\,\rm MP}=(-0.16\pm 0.20)\degree{\rm PA}/\degree$.
Likewise, neither the IP nor LFC displays a statistically significant change in the polarization degree.
However, the MP does show a small but statistically significant quadratic variation in linear-polarization degree---$p''_{L0\,\rm MP}=(0.88\pm 0.22)\%/\degree/\degree$ about its central value---$p_{L0\,\rm MP}=(23.0\pm 0.3)\%$---for a pulse-average linear polarization $\overline{p}_{L\,\rm MP}=(23.7\pm 0.3)\%$.

Our analysis of the radio Stokes data shows no strong sweep of the linear-polarization position angle. 
This lack of strong position-angle swings contrasts with the rapid swings observed in the visible band.
Current models for pulsar emission geometries do not readily account for the absence of substantial variations in both polarization degree and position angle across a pulse component (\S~\ref{s:imp}).
Thus, alternative models---e.g., dissipative magnetopheres---should be considered in modeling the radio polarization of the Crab pulsar's MP and IP.
The nearly complete polarization of the LFC suggest that it originates at a different location and via a different mechanism than do the stronger MP and IP.

Finally, the fine time resolution and high signal-to-noise ratio in the MP data led to detection of statistically significant substructure in its pulse profile.
We surmise that this substructure results from giant radio pulses occurring during the 144-minute observation.

{\bf Acknowledgments}

The Westerbork Synthesis Radio Telescope (WSRT) is operated by ASTRON, the Netherlands Institute for Radio Astronomy, with support from NWO, the Netherlands Foundation for Scientific Research. 
AS acknowledges grant DEC-2011/03/D/ST9/00656 from the Polish National Science Centre; 
BWS, a Consolidated Grant from the UK Science and Technology Facilities Council; 
AKH, NASA grants Astrophysics Theory 12-ATP12-0169 and Fermi Guest Investigator 11-FERMI11-0052; 
AJvdH, Advanced Investigator Grant 247295 (PI: R.~A.~M.~J. Wijers) from the European Research Council; 
and SLO, RFE and MCW, support by NASA's Chandra Program. 

\appendix
\section{Statistical analysis}\label{s:stat}

\subsection{Procedures}\label{ss:proc}
As Figure~\ref{f:IQUV} shows, the main pulse (MP), interpulse (IP), and low-frequency component (LFC) are well separated in the 1.38-GHz data folded on the Crab pulsar's period.
Consequently, we choose to analyze each of these three features individually, using phase ranges $(-7.2\degree,7.2\degree)$ for the MP, $(134.6\degree,156.2\degree)$ for the IP, and $(-52.1\degree,-23.3\degree)$ for the LFC, where the center of the MP defines pulse-phase angle $\varphi = 0\degree$.
We use data over the remaining phase ranges to measure the off-pulse mean and the root-of-mean-square (RMS) noise in $I$, $Q$, $U$, and $V$.
Upon measuring the off-pulse mean values for $I$, we noticed that its off-pulse value near the MP is depressed with respect to the remaining phase ranges.
Specifically, in phase ranges $(-14.4\degree,-7.2\degree)$ and $(7.2\degree,14.4\degree)$, the mean $I$ is 0.0273 ($\times1000)$ less than in other off-pulse ranges.
Taking this into account lowered $\chi^{2}$ by about 300 in fitting the $I$ pulse profile, but did not significantly alter the fitted polarization properties.
 
For convenience, we pre-process the raw data by subtracting the respective off-pulse mean value, under the assumption that the expectation values for $I$, $Q$, $U$, and $V$ are zero away from pulse features.
Furthermore, we take the RMS noise levels---0.0324, 0.0310, 0.0311, and 0.0307 (each $\times 1000$)---as estimators of the statistical standard deviations $\sigma_{I}$, $\sigma_{Q}$, $\sigma_{U}$, and $\sigma_{V}$, respectively.

In order to fit the model to the data for each pulse feature, we minimize the chi-square statistic of the combined Stokes data
\begin{eqnarray}\label{e:chi2}
\lefteqn{\chi^{2}(\varpi) = \chi_{I}^{2}(\varpi) + \chi_{Q}^{2}(\varpi)+ \chi_{U}^{2}(\varpi)+ \chi_{V}^{2}(\varpi) = } \\
                 &  & \sum_{n=1}^{N} \left[ \frac{(I_{n}-I(\varphi_{n};\varpi))^{2}}{\sigma_{I}^{2}} + \frac{(Q_{n}-Q(\varphi_{n};\varpi))^{2}}{\sigma_{Q}^{2}} + \frac{(U_{n}-U(\varphi_{n};\varpi))^{2}}{\sigma_{U}^{2}} + \frac{(V_{n}-V(\varphi_{n};\varpi))^{2}}{\sigma_{V}^{2}} \right] ,  \nonumber
\end{eqnarray}
\noindent with respect to a set $\varpi$ of $K$ model parameters, leaving $\nu = N-K$ degrees of freedom. 
We obtain the statistical uncertainty in each parameter, based upon $\Delta\chi^{2} = \chi^{2}-\chi^{2}_{\rm min}$.
To perform the $\chi^{2}$ analysis, we used the {\sl Mathematica}$^{\rm TM}$ \citep{Math9} function {\tt NonlinearModelFit}\footnote{http://reference.wolfram.com/mathematica/ref/NonlinearModelFit.html}, which finds best-fit model parameters, their errors, correlation matrix amongst them, etc.

Modeling the Stokes data requires parameterized functions for the pulse profile $I(\varphi)$, the linear-polarization fraction $p_{L}(\varphi)$, the polarization position angle $\psi(\varphi)$, and the circular-polarization fraction $p_{C}(\varphi)$ (cf.\ Equations~\ref{e:Q}, \ref{e:U}, and \ref{e:V} for $Q(\varphi)$,  $U(\varphi)$, and $V(\varphi)$, respectively).
As there is no evidence for rapid changes in polarization degree or position angle over a pulse feature (cf.\ Figures \ref{f:pol_MP} and \ref{f:pol_IP}), simple Taylor-series expansions suffice:
\begin{eqnarray}\label{e:pL2}
p_{L}(\varphi) & = & p_{L}(\varphi_{0}) + p'_{L}(\varphi_{0}) (\varphi-\varphi_{0}) + \frac{1}{2} p''_{L}(\varphi_{0}) (\varphi-\varphi_{0})^2 + \cdots \nonumber \\
  & \equiv & p_{L0} + p'_{L0} (\varphi-\varphi_{0}) + \frac{1}{2} p''_{L0} (\varphi-\varphi_{0})^2 + \cdots ;
\end{eqnarray}
\begin{eqnarray}\label{e:psi2}
\psi(\varphi) & = & \psi(\varphi_{0}) + \psi'(\varphi_{0}) (\varphi-\varphi_{0}) + \frac{1}{2} \psi''(\varphi_{0}) (\varphi-\varphi_{0})^2 + \cdots \nonumber \\
  & \equiv & \psi_{0} + \psi'_{0} (\varphi-\varphi_{0}) + \frac{1}{2} \psi''_{0} (\varphi-\varphi_{0})^2 + \cdots ;
\end{eqnarray}
\begin{eqnarray}\label{e:pC2}
p_{C}(\varphi) & = & p_{C}(\varphi_{0}) + p'_{C}(\varphi_{0}) (\varphi-\varphi_{0}) + \frac{1}{2} p''_{C}(\varphi_{0}) (\varphi-\varphi_{0})^2 + \cdots \nonumber \\
  & \equiv & p_{C0} + p'_{C0} (\varphi-\varphi_{0}) + \frac{1}{2} p''_{C0} (\varphi-\varphi_{0})^2 + \cdots .
\end{eqnarray}
To parameterize the pulse profile, we use a Gaussian (\S\ref{ss:1-G}) for each pulse feature (MP, IP, or LFC) or multiple Gaussians (\S\ref{ss:6-G}) for the MP.

\subsection{Single-Gaussian fits to the MP, the IP, and to the LFC}\label{ss:1-G}

To complete the parameterized model for the four Stokes functions, we assume a Gaussian profile:
\begin{equation}\label{e:I1}
I(\varphi) = I_{0} \exp\left(-\frac{(\varphi-\varphi_{0})^{2}}{2 \sigma^{2}_{\varphi}}\right) ,
\end{equation}
\noindent with $I_{0}$ the value of $I(\varphi)$ at pulse center, $\sigma_{\varphi}$ the Gaussian width, and $\varphi_{0}$ the phase at the pulse center.
Combining this parameterization with Equations~\ref{e:Q}, \ref{e:U}, \ref{e:V}, \ref{e:pL2}, \ref{e:psi2}, \ref{e:pC2}, the full model for the other three Stokes functions follows:
\begin{eqnarray}\label{e:Q2}
Q(\varphi) & = & I_{0} \exp\left(-\frac{(\varphi-\varphi_{0})^{2}}{2 \sigma^{2}_{\varphi}}\right) 
 [p_{L0} + p'_{L0} (\varphi-\varphi_{0})+ \frac{1}{2} p''_{L0} (\varphi-\varphi_{0})^2] \nonumber \\
 & & \times \cos(2 [\psi_{0} + \psi'_{0} (\varphi-\varphi_{0}) + \frac{1}{2} \psi''_{0} (\varphi-\varphi_{0})^2]) ;
\end{eqnarray}
\begin{eqnarray}\label{e:U2}
U(\varphi) & = & I_{0} \exp\left(-\frac{(\varphi-\varphi_{0})^{2}}{2 \sigma^{2}_{\varphi}}\right) 
 [p_{L0} + p'_{L0} (\varphi-\varphi_{0})+ \frac{1}{2} p''_{L0} (\varphi-\varphi_{0})^2] \nonumber \\
 & & \times \sin(2 [\psi_{0} + \psi'_{0} (\varphi-\varphi_{0}) + \frac{1}{2} \psi''_{0} (\varphi-\varphi_{0})^2]) ;
\end{eqnarray}
\begin{eqnarray}\label{e:V2}
V(\varphi) & = & I_{0} \exp\left(-\frac{(\varphi-\varphi_{0})^{2}}{2 \sigma^{2}_{\varphi}}\right) 
 [p_{C0} + p'_{C0} (\varphi-\varphi_{0})+ \frac{1}{2} p''_{C0} (\varphi-\varphi_{0})^2] .
\end{eqnarray}

\begin{figure}[ht]
\begin{center}
\includegraphics[angle=0,width=0.9\columnwidth]{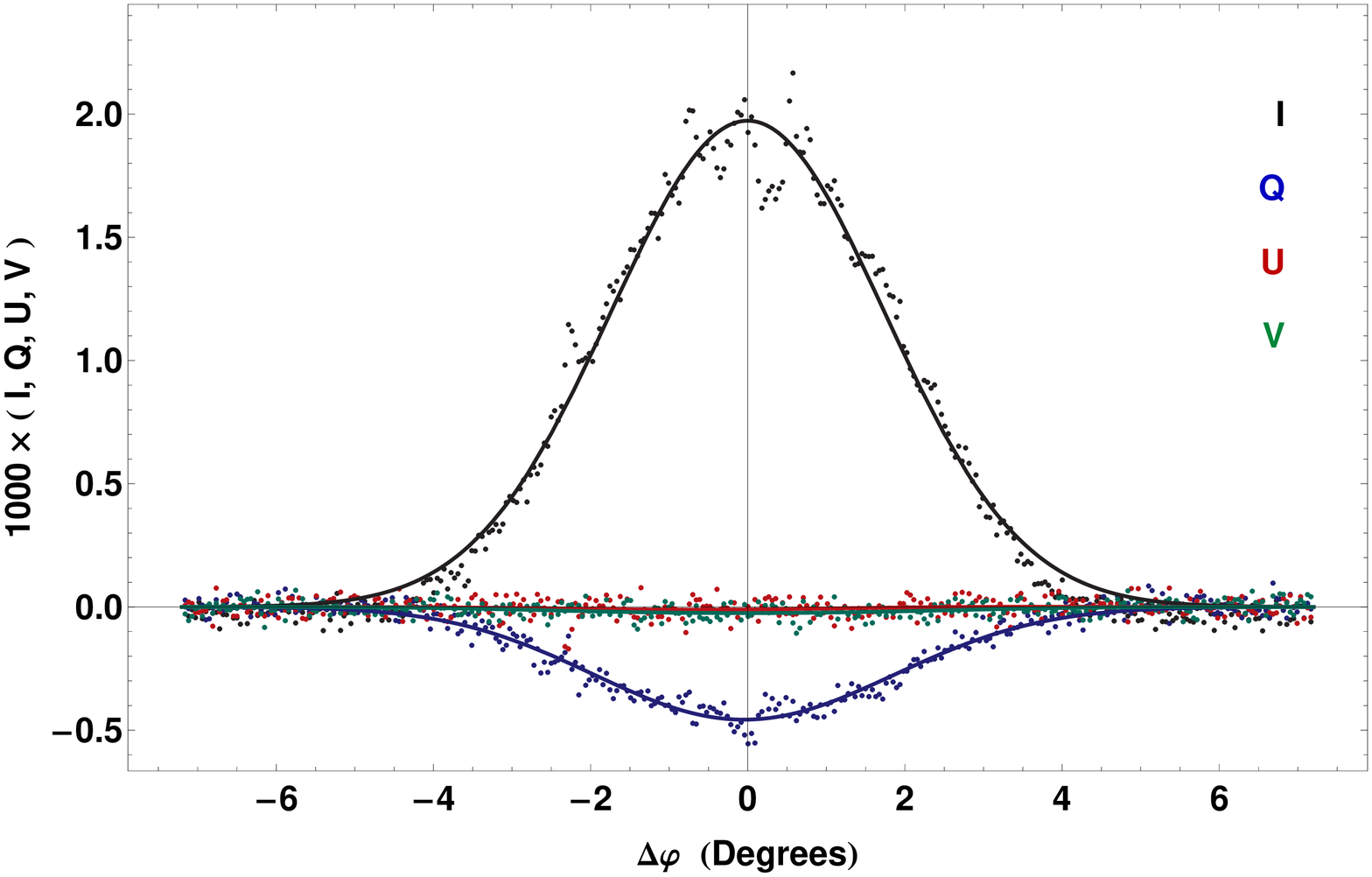}
\figcaption{Stokes data $I$, $Q$, $U$, and $V$ versus pulse phase offset $\Delta\varphi$ from the center of the main pulse (MP).
The lines represent the best-fit (minimum-$\chi^{2}$) Stokes functions for a single-Gaussian profile and up-to-second-order variations in polarization degree and in position angle.
\label{f:IQUV_MP}}
\end{center}
\end{figure}

\begin{figure}[ht]
\begin{center}
\includegraphics[angle=0,width=0.9\columnwidth]{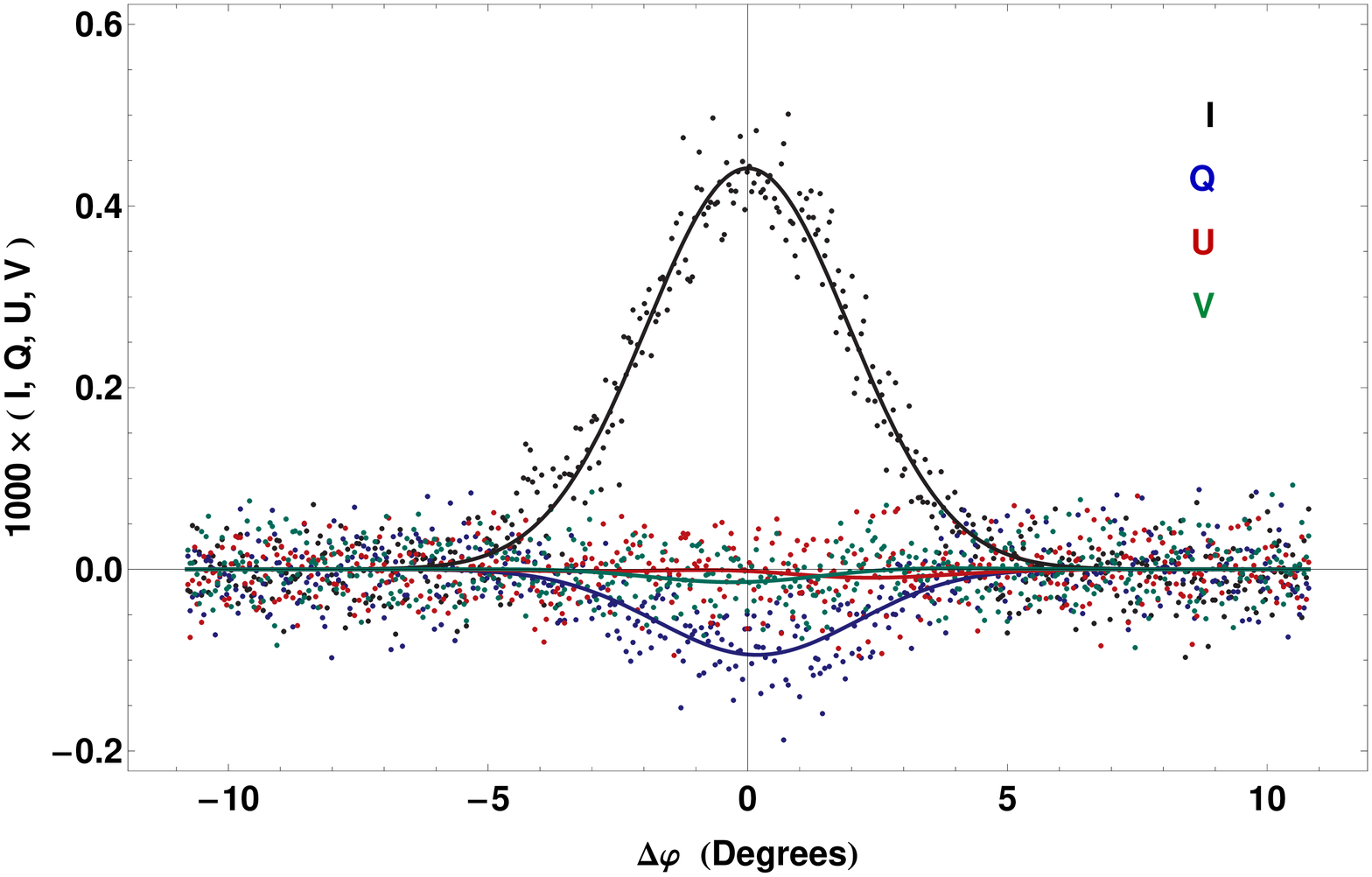}
\figcaption{Stokes data $I$, $Q$, $U$, and $V$ versus pulse phase offset $\Delta\varphi$ from the center of the inter pulse (IP).
The lines represent the best-fit (minimum-$\chi^{2}$) Stokes functions for a single-Gaussian profile and up-to-second-order variations in polarization degree and in position angle.
\label{f:IQUV_IP}}
\end{center}
\end{figure}

\begin{figure}[ht]
\begin{center}
\includegraphics[angle=0,width=0.9\columnwidth]{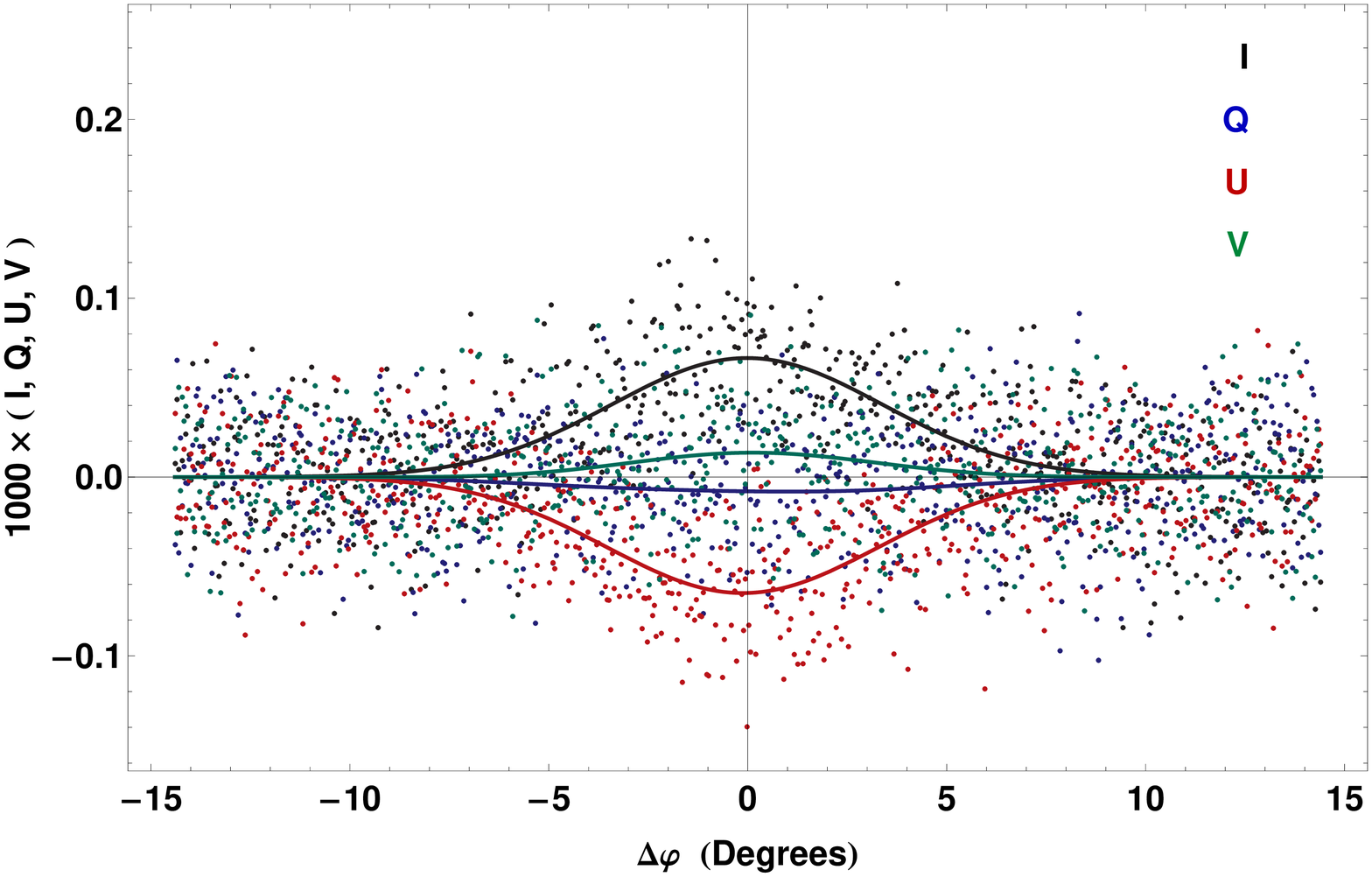}
\figcaption{Stokes data $I$, $Q$, $U$, and $V$ versus pulse phase offset $\Delta\varphi$ from the center of the low-frequency component (LFC).
The lines represent the best-fit (minimum-$\chi^{2}$) Stokes functions for a single-Gaussian profile and up-to-second-order variations in polarization degree and in position angle.
\label{f:IQUV_LFC}}
\end{center}
\end{figure}

Figures~\ref{f:IQUV_MP}, \ref{f:IQUV_IP}, and \ref{f:IQUV_LFC} display Stokes data for the MP, IP, and LFC, respectively.
The lines represent best-fit (minimum-$\chi^{2}$) Stokes functions (Equations~\ref{e:I1}, \ref{e:Q2},  \ref{e:U2}, and \ref{e:V2}) for a single-Gaussian profile $I(\varphi)$ and up-to-quadratic variations in linear-polarization degree $p_{L}(\varphi)$, in position angle $\psi(\varphi)$, and in circular-polarization degree $p_{C}(\varphi)$.
Tables~\ref{t:fit0}, \ref{t:fit1}, and \ref{t:fit2} tabulate the results of the $\chi^{2}$ analysis for a Gaussian profile and retaining polarization terms (Equations~\ref{e:Q2}, \ref{e:U2}, and \ref{e:V2}) through, zeroth, first, and second order, respectively.
For each pulse feature---MP, IP, and LFC---the tables list the minimum $\chi^2$ and degrees of freedom $\nu$ for $I$, $Q$, $U$, and $V$ data sets combined and separately, followed by best-fit estimators and (1-sigma) uncertainties for the 3 pulse-profile parameters ($I_0$, $\sigma_{\varphi}$, $\varphi_0$) and for the relevant polarization coefficients ($p_{L0}$, $p'_{L0}$, $p''_{L0}$; $\psi_0$, $\psi'_0$, $\psi''_0$; $p_{C0}$, $p'_{C0}$, $p''_{C0}$).
Note that these tables reference the pulse-phase angles ($\varphi_{0}$) and polarization position angles ($\psi_{0}$) to the MP, as we set $\varphi_{\rm MP}\equiv 0$ and were unable to obtain an absolute measurement of position angle $\psi_{\rm MP}$. 

\begin{table}[ht]
\begin{center}
\caption {Best-fit parameters for the MP, IP, and the LFC, using a simple Gaussian for each profile and no variations in polarization functions $p_{L}(\varphi)$, $\psi(\varphi)$, and $p_{C}(\varphi)$. \label{t:fit0}}
\begin{tabular}{ccccc} \\ \hline 
  Parameter            & Units        & MP                 & IP                 & LFC                \\ \hline
$\chi^2/\nu$           &              & $3081./1302$       & $2022./1962$       & $2518./2618$       \\ 
$\chi^2_I/\nu_I$       &              & $1910./324$        & $561./489$         & $577./653$         \\ 
$\chi^2_Q/\nu_Q$       &              & $460./322$         & $534./487$         & $674./651$         \\ 
$\chi^2_U/\nu_U$       &              & $441./322$         & $463./487$         & $603./651$         \\ 
$\chi^2_V/\nu_V$       &              & $269./323$         & $463./488$         & $664./652$         \\ \hline 
$I_{0}$                & $\times1000$ & $1.9894\pm 0.0046$ & $0.4414\pm 0.0044$ & $0.0668\pm 0.0031$ \\ 
$\sigma_{\varphi}$     & $\degree$    & $1.7801\pm 0.0047$ & $1.947\pm 0.022$   & $3.40\pm 0.14$     \\ 
$\varphi_{0}-\varphi_{\rm MP}$ & $\degree$ & $\equiv 0$    & $145.399\pm 0.023$ & $-37.79\pm 0.14$   \\ 
$p_{L0}$               & $\%$         & $23.67\pm 0.19$    & $21.24\pm 0.81$    & $98.3\pm 5.7$      \\ 
$\psi_{0}-\psi_{\rm MP}$ & $\degree{\rm PA}$ & $\equiv 0$  & $1.0\pm 1.1$       & $40.3\pm 1.2$      \\ 
$p_{C0}$               & $\%$         & $-1.40\pm 0.18$    & $-2.70\pm 0.78$    & $19.0\pm 4.0$      \\ \hline 
\end{tabular}
\end{center}
\end{table}

\begin{table}[ht]
\begin{center}
\caption {Best-fit parameters for the MP, the IP, and the LFC, using a simple Gaussian for each profile and up-to-linear variations in polarization functions $p_{L}(\varphi)$, $\psi(\varphi)$, and $p_{C}(\varphi)$. \label{t:fit1}}
\begin{tabular}{ccccc} \\ \hline
  Parameter           & Units         & MP                 & IP                 & LFC              \\ \hline
$\chi^2/\nu        $  &               & $3076./1299$       & $2017./1959$       & $2517./2615$     \\ 
$\chi^2_{I}/\nu_{I}$  &               & $1910./324$        & $561./489$         & $577./653$       \\ 
$\chi^2_{Q}/\nu_{Q}$  &               & $456./320$         & $532./485$         & $674./649$       \\ 
$\chi^2_{U}/\nu_{U}$  &               & $440./320$         & $462./485$         & $602./649$       \\
$\chi^2_{V}/\nu_{V}$  &               & $269./322$         & $463./487$         & $664./651$       \\ \hline
$I_{0}$               & $\times1000$  & $1.9894\pm 0.0046$ & $0.4415\pm 0.0044$ & $0.0668\pm 0.0031$ \\ 
$\sigma_\varphi$      & $\degree$     & $1.7801\pm 0.0047$ & $1.946\pm 0.022$   & $3.40\pm 0.14$   \\
$\varphi_0-\varphi_{\rm MP}$ & $\degree$ & $\equiv 0$      & $145.389\pm 0.023$ & $-37.74\pm 0.20$ \\   
$p_{L0}$              & $ \% $        & $23.67\pm 0.19$    & $21.25\pm 0.81$    & $98.3\pm 5.7$    \\
$p'_{L0}$             & $\%/\degree$  & $-0.32\pm 0.15$    & $1.09\pm 0.59$     & $- 0.9\pm 2.4$   \\
$\psi_0-\psi_{\rm MP}$ & $\degree{\rm PA}$ & $\equiv 0$    & $0.9\pm 1.1$       & $40.3\pm 1.2$    \\ 
$\psi'_0$ & $\degree{\rm PA}/\degree$ & $-0.15\pm 0.18$    & $0.91\pm 0.78$     & $-0.18\pm 0.48$  \\ 
$p_{C0}$              & $ \% $        & $-1.40\pm 0.18$    & $-2.70\pm 0.78$    & $19.0\pm 4.0$    \\
$p'_{C0}$             & $\%/\degree$  & $-0.01\pm 0.14$    & $0.38\pm 0.57$     & $0.3\pm 1.7$     \\ \hline
\end{tabular}
\end{center}
\end{table}

\begin{table}[ht]
\begin{center}
\caption {Best-fit parameters for the MP, for the IP, and for the LFC, using a simple Gaussian for each profile and up-to-quadratic variations in polarization functions $p_{L}(\varphi)$, $\psi(\varphi)$, and $p_{C}(\varphi)$. \label{t:fit2}}
\begin{tabular}{ccccc} \\ \hline 
  Parameter           & Units           & MP                 & IP                 & LFC              \\ \hline
$\chi^2/\nu        $  &                 & $3049./1296$       & $2016./1956$       & $2517./2612$     \\
$\chi^2_{I}/\nu_{I}$  &                 & $1909./324$        & $561./489$         & $577./653$       \\ 
$\chi^2_{Q}/\nu_{Q}$  &                 & $432./318$         & $531./483$         & $674./647$       \\ 
$\chi^2_{U}/\nu_{U}$  &                 & $440./318$         & $461./483$         & $603./647$       \\ 
$\chi^2_{V}/\nu_{V}$  &                 & $268./321$         & $462./486$         & $664./650$       \\ \hline
$I_{0}$               & $\times1000$    & $1.9927\pm 0.0047$ & $0.4414\pm 0.0045$ & $0.0666\pm 0.0034$ \\ 
$\sigma_\varphi$      & $\degree$       & $1.7742\pm 0.0048$ & $1.947\pm 0.023$   & $3.42\pm 0.20$   \\
$\varphi_0-\varphi_{\rm MP}$ & $\degree$ & $\equiv 0$        & $145.389\pm 0.023$ & $-37.75\pm 0.20$ \\   
$p_{L0}$              & $ \% $          & $22.99\pm 0.23$    & $21.24\pm 0.99$    & $98.1\pm 7.0$    \\
$p'_{L0}$             & $\%/\degree$    & $-0.32\pm 0.15$    & $1.03\pm 0.59$     & $-0.9\pm 2.4$    \\
$p''_{L0}$     & $\%/\degree/\degree$   & $0.86\pm 0.17$     & $-0.04\pm 0.61$    & $0.1\pm 1.4$     \\
$\psi_0-\psi_{\rm MP}$ & $\degree{\rm PA}$ & $\equiv 0$      & $-0.1\pm 1.3$      & $40.8\pm 1.4$    \\ 
$\psi'_0$ & $\degree{\rm PA}/\degree$   & $-0.16\pm 0.17$    & $0.82\pm 0.79$     & $-0.16\pm 0.48$  \\
$\psi''_0$ & $\degree{\rm PA}/\degree/\degree$ & $-0.06\pm 0.18$ & $1.07\pm 0.80$ & $-0.21\pm 0.28$  \\ 
$p_{C0}$              & $ \% $          & $-1.25\pm 0.22$    & $-3.15\pm 0.96$    & $20.5\pm 4.9$    \\
$p'_{C0}$             & $\%/\degree$    & $0.01\pm 0.15$     & $0.38\pm 0.57$     & $0.3\pm 1.7$     \\
$p''_{C0}$       & $\%/\degree/\degree$ & $-0.20\pm 0.16$    & $0.47\pm 0.59$     & $-0.49\pm 0.96$  \\ \hline
\end{tabular}
\end{center}
\end{table}

Table~\ref{t:fit1} documents that, to within statistical uncertainties, $p'_{L0} = 0$, $\psi'_{0} = 0$, and $p'_{C0} = 0$ for each of the three pulse features---MP, IP, or LFC.
Equivalently, including the three linear coefficients $p'_{L0} = 0$, $\psi'_{0} = 0$, and $p'_{C0} = 0$, does not result in a statistically significant reduction in the value of $\chi^2_{\rm min}$ (cf.\ Tables~\ref{t:fit0} and \ref{t:fit1}).
In contrast, including the quadratic parameter $p''_{L0}$ does significantly reduce the value of $\chi^2_{\rm min}$ for the MP (cf.\ Table~\ref{t:fit2} with Table~\ref{t:fit1} or \ref{t:fit0}), but not for the IP nor for the LFC.

\subsection{Comparison of model fits to MP}\label{ss:6-G}

Table~\ref{t:fit0} shows that a single-Gaussian profile and constant polarization degree and position angle provide a statistically adequate fit to the Stokes data for the IP and for the LFC.
However, the simple model does not provide a statistically adequate fit to the Stokes data for the MP, at least in part due to the higher signal-to-noise ratio in the MP Stokes data.
Consequently, we here investigate more complicated models in order to improve the goodness of the $\chi^2$ fits to the MP Stokes data.
In particular, we investigate using a multi-Gaussian function for the MP pulse profile. 
Table~\ref{t:fit6} lists the minimum $\chi^2$ and degrees of freedom $\nu$ for $I$, $Q$, $U$, and $V$ data sets combined and separately, followed by best-fit estimators and (1-sigma) uncertainties for the 9 polarization coefficients ($p_{L0}$, $p'_{L0}$, $p''_{L0}$; $\psi_0$, $\psi'_0$, $\psi''_0$; $p_{C0}$, $p'_{C0}$, $p''_{C0}$) of the Taylor expansion through second order.

\begin{table}[ht]
\begin{center}
\caption {Comparison of results of fitting the main pulse (MP) profile with a simple Gaussian, with a multi-Gaussian, and with a simple Gaussian after adjusting weightings. 
The models retain up-to-quadratic variations in the polarization functions $p_{L}(\varphi)$, $\psi(\varphi)$, and $p_{C}(\varphi)$.\label{t:fit6}}
\begin{tabular}{ccccc} \\ \hline 
  Parameter          & Units              & 1-Gaussian        & 6-Gaussian        & 1-Gaussian (Adj.) \\ \hline
$\chi^2/\nu$         &                    & $3049./1296$      & $1823./1281$      & $1281./1296$      \\ 
$\chi^2_{I}/\nu_{I}$ &                    & $1909./324$       & $688./309$        & $324./324$        \\ 
$\chi^2_{Q}/\nu_{Q}$ &                    & $432./318$        & $430./303$        & $318./318$        \\ 
$\chi^2_{U}/\nu_{U}$ &                    & $440./318$        & $438./303$        & $318./318$        \\
$\chi^2_{V}/\nu_{V}$ &                    & $268./321$        & $267./306$        & $321./321$        \\ \hline
$p_{L0}$             & $ \% $             & $22.99\pm 0.23$   & $22.91\pm 0.24$   & $22.98\pm 0.30$   \\
$p'_{L0}$            & $\%/\degree$       & $-0.32\pm 0.15$   & $-0.29\pm 0.15$   & $-0.31\pm 0.19$   \\
$p''_{L0}$        & $\%/\degree/\degree$  & $0.86\pm 0.17$    & $0.89\pm 0.19$    & $0.88\pm 0.22$    \\
$\psi_0$            & $\degree{\rm PA}$   & $-89.34\pm 0.27$  & $-89.38\pm 0.29$  & $-89.34\pm 0.32$  \\ 
$\psi'_0$ & $\degree{\rm PA}/\degree$     & $-0.16\pm 0.17$   & $-0.19\pm 0.17$   & $-0.16\pm 0.20$   \\
$\psi''_0$ & $\degree{\rm PA}/\degree/\degree$ & $-0.06\pm 0.18$ & $0.05\pm 0.20$ & $-0.06\pm 0.21$   \\ 
$p_{C0}$             & $ \% $             & $-1.25\pm 0.22$   & $-1.27\pm 0.23$   & $-1.25\pm 0.20$   \\
$p'_{C0}$            & $\%/\degree$       & $-0.01\pm 0.15$   & $-0.02\pm 0.14$   & $-0.01\pm 0.13$   \\
$p''_{C0}$        & $\%/\degree/\degree$  & $-0.20\pm 0.16$   & $-0.18\pm 0.18$   & $-0.20\pm 0.15$   \\ \hline
\end{tabular}
\end{center}
\end{table}

Comparison of the column ``MP'' in Table~\ref{t:fit1} with that in Table~\ref{t:fit2} (or, equivalently, with the column ``1-Gaussian'' in Table~\ref{t:fit6}) finds that inclusion of the three quadratic polarization coefficients---especially $p''_{L0}$---reduces $\chi^2_{Q}$ by 42 (from 473 to 431). 
While $\psi''_{0} = 0$ and $p''_{C0} = 0$ within statistical uncertainties, $p''_{L0} \approx (0.9\pm 0.2) \%/\degree/\degree$ is statistically significant but small.

The main cause of the poor fit of the 1-Gaussian model to the MP data, however, has nothing to do with polarization.
Figure~\ref{f:IQUV_MP_2+4} illustrates that, for the fine time resolution and the high signal-to-noise ratio of the MP data, substructure in the pulse profile is quite evident.
Using a 6-Gaussian (2 broad and 4 narrow) profile for $I(\varphi)$ substantially improves the fit.
Comparing the column ``6-Gaussian'' with ``1-Gaussian'' in Table~\ref{t:fit6} finds that inclusion of $15=5\times 3$ additional (Gaussian) parameters reduces $\chi^2_{I}$ by 1221 (from 1909 to 688).
Even so, the fit to the Stokes data is not formally acceptable.

It is important to note that the best-fit expectation values and uncertainties for the polarization coefficients ($p_{L0}$, $p'_{L0}$, $p''_{L0}$; $\psi_0$, $\psi'_0$, $\psi''_0$; $p_{C0}$, $p'_{C0}$, $p''_{C0}$) are rather insensitive to details of the pulse profile.
Thus, we compensate for fine substructure in the pulse profile by increasing the estimators for the measurement standard deviations until a statistically acceptable fit is achieved.
That is, we adjust $\sigma_{I}$, $\sigma_{Q}$, $\sigma_{U}$, and $\sigma_{V}$ until (Eq.~\ref{e:chi2}) $\chi_{I}^{2}/\nu_{I}$, $\chi_{Q}^{2}/\nu_{Q}$, $\chi_{U}^{2}/\nu_{U}$, and $\chi_{V}^{2}/\nu_{V}$, respectively, are close to unity.
The column ``1-Gaussian (Adj.)'' in Table~\ref{t:fit6} shows the best-fit polarization parameters for a single-Gaussian profile, with weightings adjusted as described.
The only noticeable effect of this adjustment upon the best-fit polarization parameters is a small change---typically an increase---in their uncertainties.
The uncertainties quoted in Table~\ref{t:fit} (\S\ref{s:res}) are the typically more conservative values obtained using the single-Gaussian profiles and adjusted weightings.


\end{document}